\documentclass[floats,floatfix,showpacs,amssymb,prd,twocolumn,superscriptaddress,nofootinbib,nolongbibliography,reprint]{revtex4-1}

\usepackage{amssymb,amsmath,verbatim,mathtools,needspace,enumitem,etoolbox,graphicx,physics,microtype,afterpage,xspace,tabularx,lmodern,multirow}
\usepackage{gensymb}
\usepackage[dvipsnames, usenames]{xcolor}
\definecolor{linkcolor}{rgb}{0.0,0.3,0.5}
\usepackage[unicode, colorlinks=true, linkcolor=linkcolor, citecolor=linkcolor, filecolor=linkcolor, urlcolor=linkcolor, linktocpage, breaklinks]{hyperref}
\usepackage[all]{hypcap}
\usepackage[T1]{fontenc}
\usepackage[utf8]{inputenc}
\usepackage[usenames,dvipsnames]{xcolor}
\hypersetup{colorlinks=true,citecolor=romared,linkcolor=romared,urlcolor=romared}

\definecolor{romared}{RGB}{142,0,28}

\newcommand{\be}{\begin{equation}}
\newcommand{\ee}{\end{equation}}

\def\be{\begin{equation}}
\def\ee{\end{equation}}
\newcommand{\beq}{\begin{eqnarray}}
\newcommand{\eeq}{\end{eqnarray}}
\def\ba{\begin{align}}
\def\ea{\end{align}}

\usepackage{aas_macros}
\usepackage{makecell}
\usepackage{soul}

\def\be{\begin{equation}}
\def\ee{\end{equation}}

\begin{document}
\title{Divergences in gravitational-wave emission and absorption\\ from extreme mass ratio binaries}

\author{Enrico Barausse}
\affiliation{SISSA, Via Bonomea 265, 34136 Trieste, Italy and INFN Sezione di Trieste}
\affiliation{IFPU - Institute for Fundamental Physics of the Universe, Via Beirut 2, 34014 Trieste, Italy}

\author{Emanuele Berti}
\affiliation{Department of Physics and Astronomy, Johns Hopkins University, 3400 N. Charles Street, Baltimore, MD 21218, US}

\author{Vitor Cardoso}
\affiliation{CENTRA, Departamento de F\'{\i}sica, Instituto Superior T\'ecnico -- IST, Universidade de Lisboa -- UL,
Avenida Rovisco Pais 1, 1049 Lisboa, Portugal}

\author{Scott A.~Hughes}
\affiliation{Department  of  Physics  and  MIT  Kavli  Institute, Massachusetts Institute of Technology, Cambridge, MA 02139, USA}

\author{Gaurav Khanna}
\affiliation{Department of Physics, University of Rhode Island, Kingston, RI 02881}
\affiliation{Center for Scientific Computing \& Visualization Research and Physics Department, University of Massachusetts, Dartmouth, MA 02747}

\begin{abstract}
  A powerful technique to calculate gravitational radiation from binary systems involves a perturbative expansion: if the masses of the two bodies are very different, the ``small'' body is treated as a point particle of mass $m_p$ moving in the gravitational field generated by the large mass $M$, and one keeps only linear terms in the small mass ratio $m_p/M$. This technique usually yields finite answers, which are often in good agreement with fully nonlinear numerical relativity results, even when extrapolated to nearly comparable mass ratios. Here we study two situations in which the point-particle approximation yields a divergent result: the instantaneous flux emitted by a small body as it orbits the light ring of a black hole, and the total energy absorbed by the horizon when a small body plunges into a black hole.  By integrating the Teukolsky (or Zerilli/Regge-Wheeler) equations in the frequency and time domains we show that both of these quantities diverge.  We find that these divergences are an artifact of the point-particle idealization, and are able to interpret and regularize this behavior by introducing a finite size for the point particle.  These divergences do not play a role in black-hole imaging, e.g. by the Event Horizon Telescope.
\end{abstract}

\maketitle

\section{Introduction}
The problem of motion and radiation emission in general relativity is notoriously difficult, and different perturbative frameworks have been developed over the years to handle it. In particular, when dealing with very asymmetric compact binaries, a perturbative expansion around the spacetime of the most massive body is generally applicable. In this approach, the lighter object -- typically, a neutron star or a stellar-mass black hole (BH) moving around or plunging into a supermassive BH -- is modeled as a pointlike particle.
This procedure was used extensively to study the gravitational-wave (GW) emission from compact objects around Schwarzschild and Kerr BHs (see e.g.~\cite{Davis:1971gg,Davis:1972ud,Detweiler:1978ge,Nakamura:1987zz, Mino:1996nk,Quinn:1996am,Sundararajan:2007jg,Poisson:2011nh,Barack:2018yvs}). 
This problem is interesting for a number of applications, including the modeling of waveforms from the inspiral or plunge of stellar-mass compact objects into  massive BHs. The latter are expected to have masses ranging at least between $10^5 M_\odot$ and $10^9 M_\odot$ and be present at the centers of most galaxies in the local low-redshift universe~\cite{1984ApJ...278...11G,Kormendy:1995er,2011Natur.470...66R,Reines:2013pia, Baldassare2019}, including our own Milky Way~\cite{1999ApJ...524..816R,Schodel:2002vg,Reid_2003,Gillessen:2008qv} and M87~\cite{Akiyama:2019cqa}. These extreme mass-ratio binaries are expected to be a major class of GW sources for the Laser Interferometer Space Antenna (LISA)~\cite{Audley:2017drz}, a joint ESA-NASA space-borne GW detector to be launched in the next decade: see e.g. Ref.~\cite{Babak:2017tow}. 

Technically, and at leading order in the mass ratio, the plunging/inspiraling particle is assumed to be pointlike and to follow a geodesic (see e.g. \cite{Poisson:2011nh} and references therein for a discussion). The radiated GW energy or fluxes at infinity are normally found to be finite: to leading order they depend only on the binary masses, and not on their internal composition.

In this paper we wish to revisit and better understand two noteworthy exceptions to this general property. They concern the total GW energy radiated {\it into the horizon} by a plunging particle~\cite{Davis:1972ud}, and the instantaneous GW energy radiated {\it into the horizon and at infinity} by a relativistic particle orbiting the light ring (i.e., the circular photon orbit)~\cite{Misner:1972jf,PhysRevD.7.1002,ruffini}. These two exceptions have received little attention, partly because of the hope that higher orders in perturbation theory might perhaps regularize the divergences, but most of all for their perceived lack of observational significance.
Indeed, the energy absorbed by the horizon is expected to only have a subtle influence on the inspiral of an orbiting body, thanks to its duality with the notion of ``tidal heating'' of a BH \cite{Datta:2019epe}. Furthermore, astrophysically relevant orbits for which these effects may be observable become unstable at the spacetime's innermost stable circular orbit (ISCO), long before reaching the light ring.  For these reasons, light-ring orbits are usually considered of mostly academic interest (but see Refs.~\cite{1968ApJ...151..659A,1972ApJ...172L..95G,Ferrari:1984zz,Schutz:1985km,Cardoso:2008bp,Yang:2012he,Cardoso:2019rvt,Akiyama:2019cqa,Volkel:2020daa,Yang:2021zqy,Cardoso:2021sip} for the connection between these orbits, BH dynamical relaxation and BH shadows).

However, early papers on this topic did consider these cases and noted their anomalous behavior. For example, studies of GW emission from unstable timelike circular orbits -- including orbits approaching asymptotically the circular photon orbit -- produced striking results.
In the early Seventies, analytic work in the Wentzel–Kramers–Brillouin (WKB) approximation suggested that (instantaneous) gravitational fluxes could become very large close to the circular photon orbit, and that they should formally \textit{diverge} as that orbit is approached~\cite{Misner:1972jf,PhysRevD.7.1002,ruffini,Davis:1972dm}.
The predicted divergence is logarithmic in the multipole number $\ell$ of the perturbations (i.e. in the limit of large $\ell$, these WKB results suggest that the contribution to the flux from multipole number $\ell$ scales with $1/\ell$). The divergence may be regularized by taking into account the finite size of the photon wave-packet, but it is nevertheless quite surprising. 
Numerical results, again in the seventies, confirmed the overall conclusion that the fluxes become very large when the photon orbit is approached, but seem to suggest a slightly faster falloff with $\ell$ (cf. Fig.~2 in Ref.~\cite{Davis:1972dm}), which would seem to make the fluxes at the photon orbit formally finite. 
More recently, similar seemingly logarithmic divergences have been found in the conservative gravitational self-force~\cite{Akcay:2012ea}.

Early work also found that the {\it total} energy  crossing the horizon when a small particle plunges into a Schwarzschild BH  diverges~\cite{Davis:1972ud}. This is again a high-frequency, large-$\ell$ divergence, which could presumably be cured by truncating the $\ell$-sum to take into account the finite size of the infalling body.

Here we revisit the problem of calculating the gravitational fluxes from relativistic particles on circular orbits in a BH spacetime, as well as the total energy radiated by a particle plunging into the horizon of a BH. These are theoretically interesting situations: the corresponding divergences require some regularization mechanism that does not seem to be needed in other similar processes. We will indeed show that finite-size effects suffice to regularize these quantities.
Throughout this paper we use geometrical units ($G=c=1$).

\section{The numerical set-up\label{sec:setup}}

\subsection{The Teukolsky equation for relativistic sources}

The Teukolsky master equation~\cite{Teukolsky:1973ha}, describing the perturbations induced by a particle of mass $m_p\ll M$, reads
\begin{eqnarray}
\label{eqn:teuk}
&&
-\left[\frac{(r^2 + a^2)^2 }{\Delta}-a^2\sin^2\theta\right]
        \partial_{tt}\Psi
-\frac{4 M a r}{\Delta}
        \partial_{t\phi}\Psi \nonumber \\
&&- 2s\left[r-\frac{M(r^2-a^2)}{\Delta}+ia\cos\theta\right]
        \partial_t\Psi\nonumber\\  
&&
+\,\Delta^{-s}\partial_r\left(\Delta^{s+1}\partial_r\Psi\right)
+\frac{1}{\sin\theta}\partial_\theta
\left(\sin\theta\partial_\theta\Psi\right)+\nonumber\\
&& \left[\frac{1}{\sin^2\theta}-\frac{a^2}{\Delta}\right] 
\partial_{\phi\phi}\Psi +\, 2s \left[\frac{a (r-M)}{\Delta} 
+ \frac{i \cos\theta}{\sin^2\theta}\right] \partial_\phi\Psi  \nonumber\\
&&- \left(s^2 \cot^2\theta - s \right) \Psi = -4\pi\left(r^2+a^2\cos^2\theta\right)T   ,
\end{eqnarray}
where $M$ is the mass of the BH, $a=J/M$ its angular momentum per unit mass, $\Delta = r^2 - 2 M r + a^2$, and $s$ is the ``spin weight'' of the field. The $s = -2$ version of this equation describes the radiative degrees of freedom of the gravitational field  in the radiation zone, and is directly related to the Weyl curvature scalar $\psi_4$ by $\Psi = (r - ia\cos\theta)^4\psi_4$.

The source term $T$ of the Teukolsky equation~\eqref{eqn:teuk} is constructed by projecting the point-particle energy-momentum tensor onto the Kinnersley tetrad. The energy-momentum tensor of a point particle exhibits a divergence at the light ring, but this divergence can be factored out, as shown in Appendix A of Ref.~\cite{taracchini2013modeling}. This is accomplished by simply rewriting the point-particle energy-momentum tensor and the geodesic equation in terms of the four-momentum per unit orbital energy, i.e. $\hat {p^\mu} = {p^\mu}/E$, which is always finite. In this way the divergence is isolated in the energy term $E$ alone (for circular orbits), enabling a numerical study of the behavior of the fluxes at the light ring in the zero-mass (i.e. photon) limit. 

\subsection{Time and frequency domain integrators\label{subsec:num_method}}

With the  modifications of the source-term described above, Eq.~\eqref{eqn:teuk} can be solved in
either the frequency domain or  the time domain. The former is particularly convenient for
pointlike particles and bound orbits, while the latter is better suited for extended objects and/or unbound orbits.

In the frequency domain, we follow the approach of Refs.~\cite{Hughes:1999bq,Hughes:2001jr,Drasco:2005kz} and solve the Teukolsky equation by separation of variables, i.e. by converting it into a pair of ordinary differential equations for the radial and angular eigenfunctions. The angular equation yields the $s=-2$ spin-weighted spheroidal harmonics, which are used to decompose the source term. The radial equation is then solved by finding homogeneous solutions numerically, and constructing the Green's function out of their Wronskian. To solve the radial Teukolsky equation, we use two codes that are based on the method developed by Mano, Suzuki, and Takasugi \cite{Mano:1996vt}.  One is {\sc Gremlin}, a {\tt C++} language code primarily developed by author Hughes and recently described in Ref.\ \cite{Hughes:2021exa}. An open-source version of this code, specialized to circular and equatorial orbits (adequate for the purposes of this study) is available through the Black Hole Perturbation Toolkit \cite{BHPToolkit} (hereafter ``the Toolkit'').  The other is a Teukolsky solver written in Mathematica \cite{mathematica}, also available through the Toolkit \cite{BHPToolkit}.

{\sc Gremlin} is fast, but is mostly limited to double-precision accuracy analyses.  As such, we primarily use it in this study to focus on multipoles with spheroidal harmonic index $\ell \lesssim 75$.  The Toolkit's Mathematica-based solver allows for the use of arbitrary-precision arithmetic, extending the range of multipoles we can to study to essentially arbitrary order, though requiring far more computing time.  Using {\sc Gremlin}, computing all the multipolar contributions over the range $2 \le \ell \lesssim 75$, $-\ell \le m \le \ell$ for several dozen orbits from the light-ring to the ISCO requires several hours on a single CPU.  By contrast, a single multipole for $\ell\sim$ several hundred may require several hours of computing time using the Mathematica-based Teukolsky solver.  As such, our high-$\ell$ data are much more sparsely sampled than our low-$\ell$ data.

In the time domain, we use an approach similar to that of Refs.~\cite{Zenginoglu:2011zz, Sundararajan:2008zm,Sundararajan:2007jg,McKennon:2012iq}. First, the Teukolsky equation is rewritten using compactified hyperboloidal coordinates that allow for GW extraction directly at null infinity, while also solving the issue of unphysical reflections from the artificial boundary of the finite computational grid. Next, we can leverage the axisymmetry of the background Kerr spacetime to separate out the dependence on the $\phi$ coordinate, thus obtaining a system of $(2+1)$ dimensional partial differential equations (PDEs). This system is then recast into first-order, hyperbolic PDE form to make it well suited for stable numerical computations. A high-order, time-explicit, WENO finite-difference numerical evolution scheme is implemented in a high-performing, OpenCL/CUDA-based GPGPU code. Additional details may be found in Refs.~\cite{Zenginoglu:2011zz, Sundararajan:2008zm,Sundararajan:2007jg,McKennon:2012iq}, although those refer to an obsolete second-order, Lax-Wendroff numerical scheme. Details on the high-order WENO implementation may be found in Ref.~\cite{Field:2021}. Numerical errors from such computations are typically well below $1\,\%$~\cite{McKennon:2012iq}.

\section{Point particle fluxes \label{sec:Divergences}}
In this section, we compute the energy flux by a particle in circular orbit at and near the light ring (Sec.~\ref{sec:LR}), as well as the energy absorbed at the horizon by a particle plunging radially (Sec.~\ref{sec:hor}).
In both cases the particle is modeled as a pointlike object of rest mass $m_p\ll M$.

\begin{figure}[t]
\includegraphics[width=0.5\textwidth]{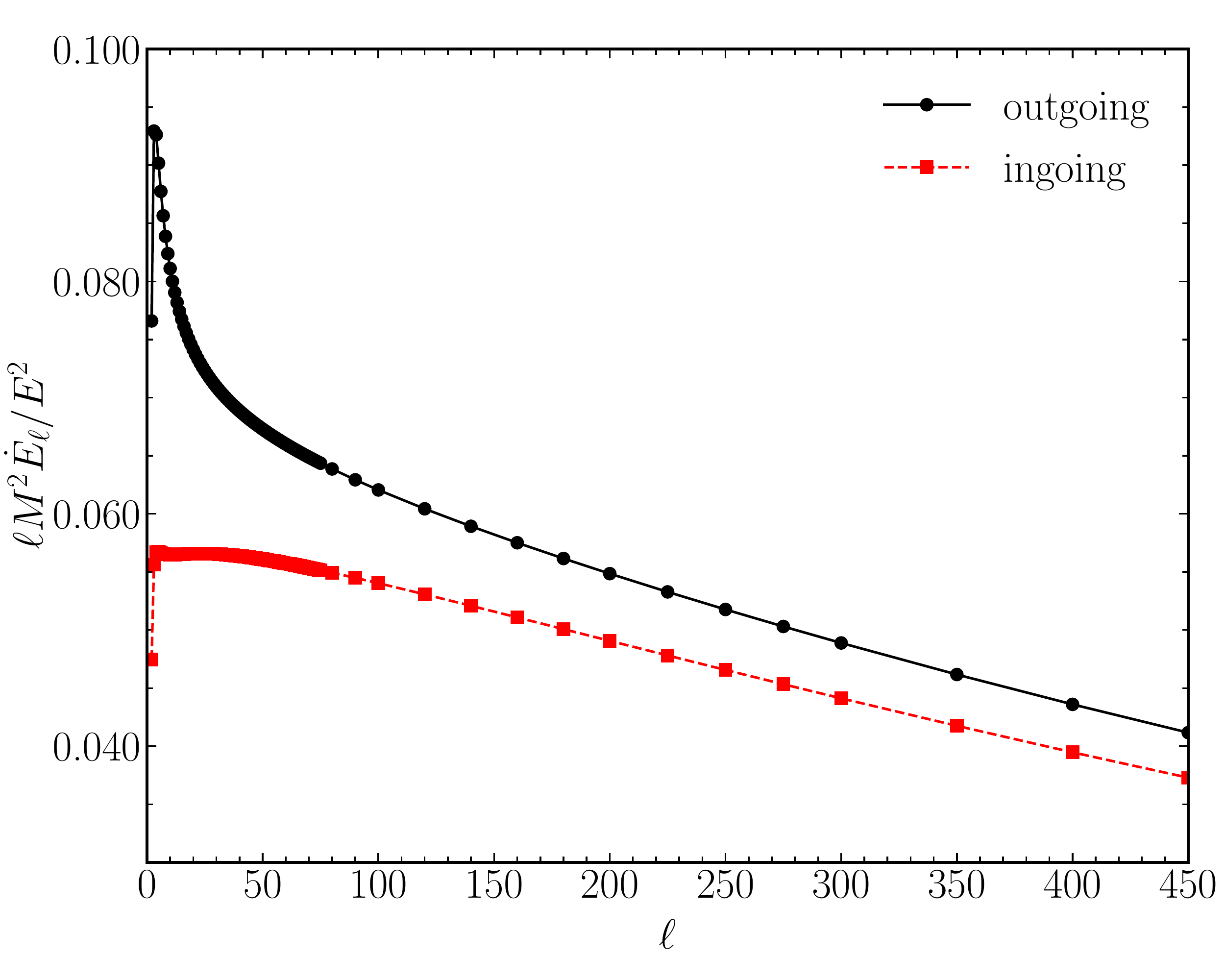}
\caption{Instantaneous gravitational-energy fluxes $\dot{E}_{\rm \ell}$
in the modes with multipole number $\ell$, for a point particle around a Schwarzschild BH on a circular orbit of radius $r=3.001 M$. We show both the gravitational energy flux at  infinity and at the horizon. We normalize the results by the square of the Killing energy $E^2$.
}
\label{fig:nearLR0}
\end{figure}

\begin{figure}[t]
\includegraphics[width=0.5\textwidth]{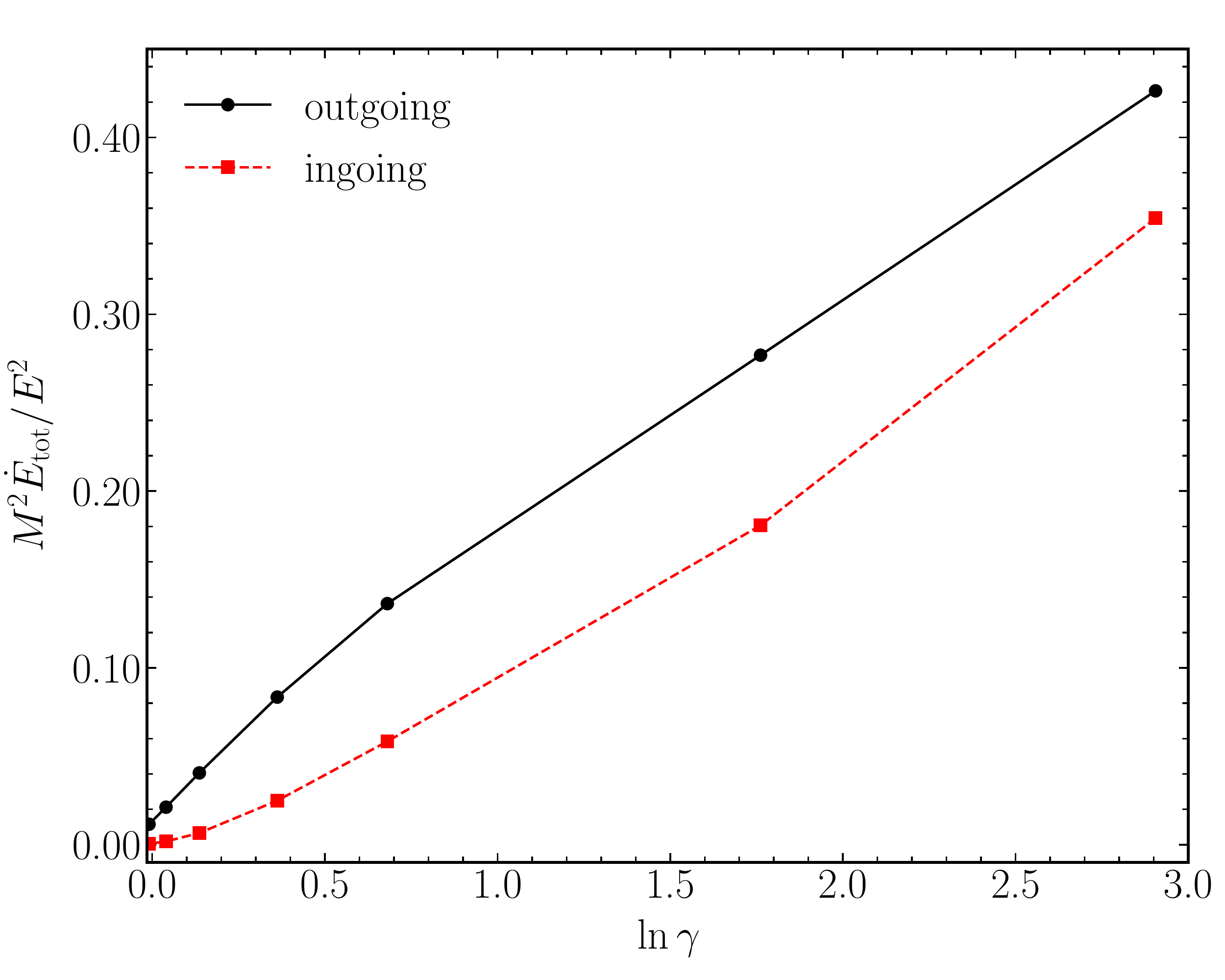}
\caption{Total instantaneous gravitational-energy fluxes $\dot{E}_{\rm tot}$ of a point particle around a Schwarzschild BH as a function of the relativistic boost factor $\gamma\equiv E/m_p$. We show the gravitational energy flux normalized by the square of the Killing energy $E^2$, both at  infinity and at the horizon.}
\label{fig:nearLR}
\end{figure}

\begin{figure*}
    \includegraphics[width=0.485\textwidth]{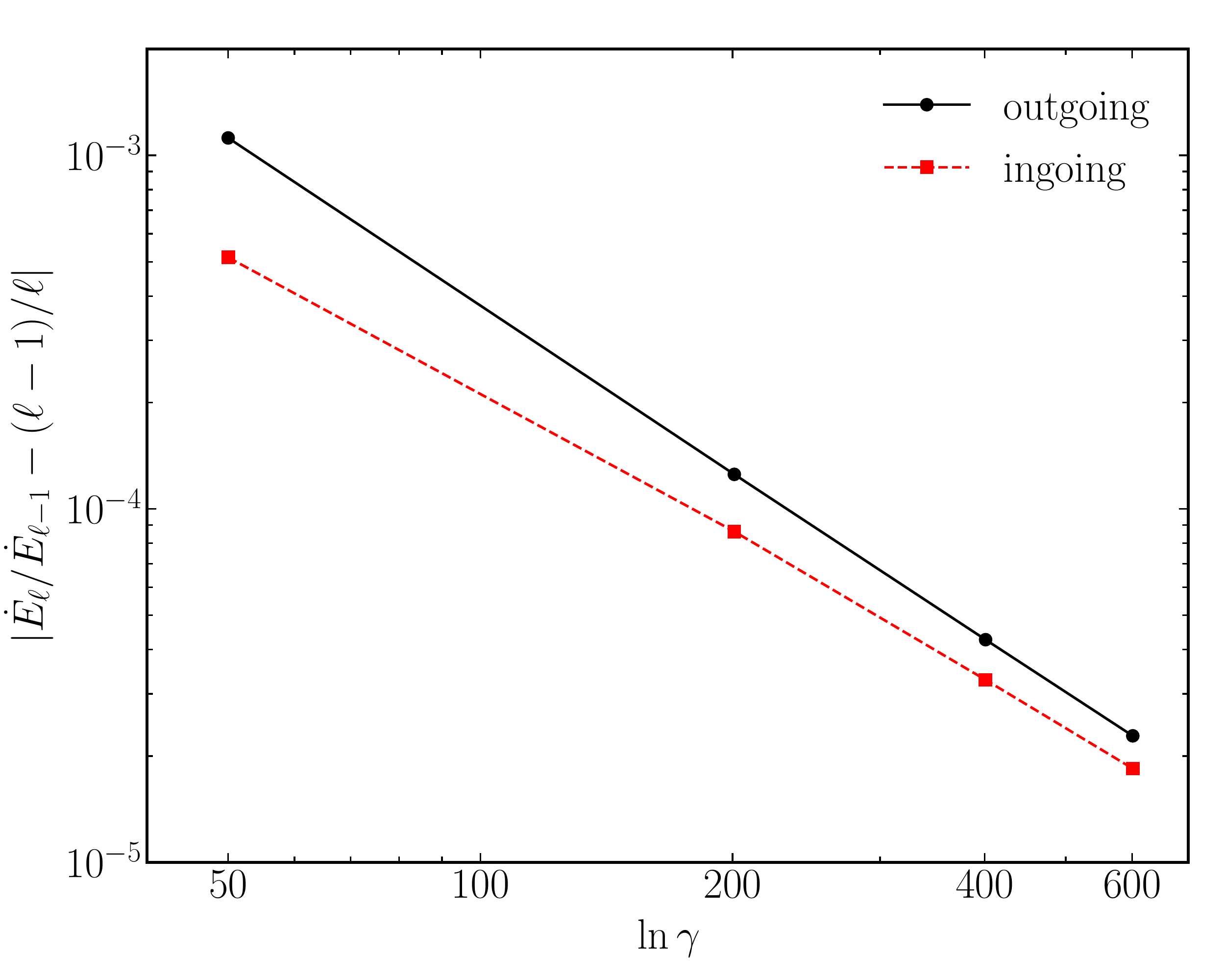}
     \includegraphics[width=0.485\textwidth]{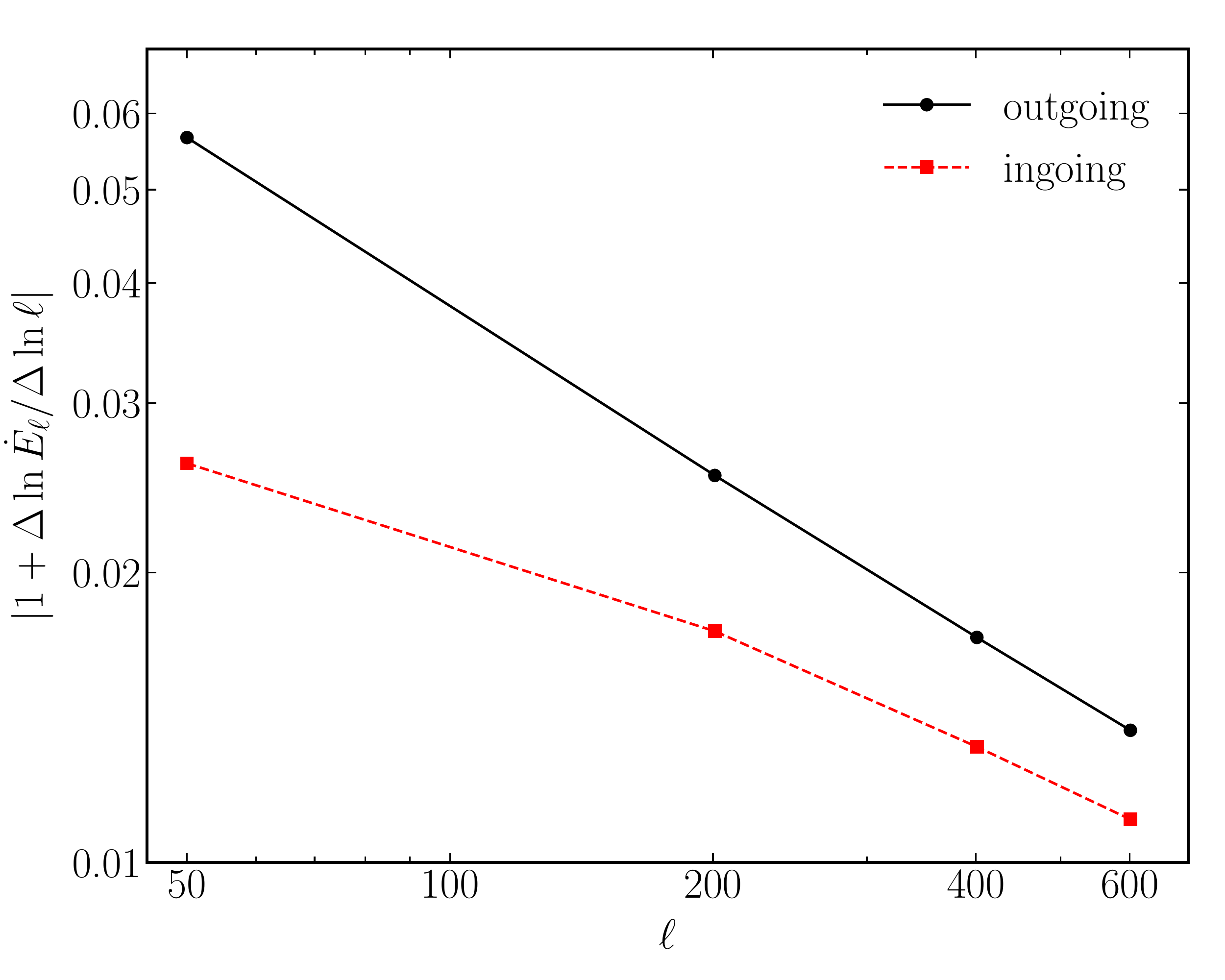}
  \caption{Scaling of the GW fluxes emitted by a (massless) particle at the circular null orbit of a Schwarzschild BH, as a function of the multipole number $\ell$. The contribution of a given multipole number is denoted by $\dot{E}_\ell$, while $\Delta \ln\dot{E}_\ell\equiv \ln\dot{E}_{\ell+1}-\ln\dot{E}_\ell$ and $\Delta \ln\ell\equiv \ln(\ell+1)-\ln\ell$. Both panels show that in the limit $\ell\to\infty$, $\dot{E}_\ell$ approaches a $1/\ell$ scaling, and therefore that the total instantaneous fluxes are logarithmically divergent.\label{fig:ratio}}
\end{figure*}

\subsection{Energy fluxes at the light ring}
\label{sec:LR}

With the formalism outlined above, and in particular with our frequency-domain Teukolsky code, we can compute GW emission from circular orbits (both timelike and null) in the Kerr spacetime. To investigate whether the fluxes from null circular orbits are large but formally finite (as suggested by the numerical results in Fig.~2 of Ref.~\cite{Davis:1972dm}; see however Ref.~\cite{Detweiler:1978ge}) or diverge logarithmically (as suggested by WKB calculations~\cite{Misner:1972jf,PhysRevD.7.1002,ruffini}), and to explore how these fluxes compare to those from neighboring but \textit{timelike} orbits, we consider sequences of timelike circular orbits approaching the light ring.  As described in Sec.~\ref{subsec:num_method} above, by combining output from {\sc Gremlin} with the Toolkit's~\cite{BHPToolkit} Mathematica-based solver, we densely sample data for multipoles with $\ell \lesssim 75$, but can examine the contributions at multipole orders up to $\ell\sim \mbox{several hundred}$.

As an example, in Fig.~\ref{fig:nearLR0} we show the fluxes emitted by a particle on a circular orbit with $r=3.001 M$ in a Schwarzschild spacetime. The energy flux (both ingoing and outgoing) in modes of multipole number $\ell$, $\dot{E}_{\ell}$, is normalized by the square of the particle's Killing energy $E$ (i.e., the energy measured by an observer at spatial infinity) and multiplied by $\ell$ itself.  Note that the rescaling of $\dot{E}_{\rm tot}$ by $E^2$ is inspired by the known scaling $\dot{E}_{\rm tot}\propto m_p^2$ that holds for orbits with $E\sim m_p$, and by the physical expectation that in the ultrarelativistic limit $E\gg m_p$ the rest mass of the particle should be unimportant (in general relativity what gravitates should be the energy and not the rest mass: see e.g. Refs.~\cite{Barausse:2010ka,Barausse:2011vx,Gundlach:2012aj,Sperhake:2012me}).  The fluxes, after an initial ``transient'' at very low $\ell$, are consistent with an initial fall-off $\propto 1/\ell$, followed by an exponential cut-off. 

By analyzing the fluxes from a sequence of circular timelike orbits approaching the light ring, at large $\ell$ and large $\gamma^2\ll \ell$ we find the following universal behavior:
\be
\dot{E}_\ell=\frac{c_0}{\ell}e^{-c_1\ell/\gamma^2}\,\frac{E^2}{M^2}\,,\quad \ell/\gamma^2 \to \infty\,,\label{rel_particles}
\ee
where $\gamma\equiv E/m_p$ is the particle's relativistic boost factor. For particles on circular orbits close to the light ring, with $r=3M(1+\delta)$, we have $\delta^{-1} \sim 9\gamma^2$. 
The scaling above, with $c_0\sim0.07\pm 0.01$, applies both to fluxes into the horizon and to fluxes at infinity, as long as $\delta \ll 1$. The constant $c_1\sim 0.42\pm 0.02$ in this limit. These constants were obtained by fitting the numerical data with $\ell \lesssim 400$ to Eq.~\eqref{rel_particles}, and using only Lorentz factors $\gamma\lesssim 18$. Note that a WKB analysis predicts $c_1=\pi/6\sim 0.52$~\cite{Misner:1972jf,Chitre:1972fv}.

We can then use this asymptotic behavior to estimate the total flux: we calculate numerically the fluxes up to the maximum multipole treatable with our numerical setup, and then use the expression above to estimate the remainder.  Our results are summarized in Fig.~\ref{fig:nearLR},
which shows the instantaneous energy fluxes $\dot{E}_{\rm tot}$ (summed over all multipole numbers $\ell$ and normalized by the square of the particle's Killing energy $E$) as functions of $\ln \gamma$. Note the apparent logarithmic divergence of both (rescaled) fluxes as we approach the null circular orbit, which corresponds to $\gamma\to\infty$.
Our results suggest that
\be\label{fluxTL}
\frac{\dot{E}_{\rm tot} M^2}{E^2} \sim k_0+ k_1\ln \gamma\,.
\ee
Here, $k_1=0.12\pm0.01$ for both ingoing and outgoing fluxes (again when $\delta \ll 1$, i.e. $\gamma\gg 1$).
A similar logarithmic divergence in the fluxes was observed for plunging orbits with arbitrary energy and angular momentum in a Schwarzschild background when the critical angular momentum corresponding to the light ring is approached ``from below''~\cite{Berti:2010ce}.

Let us now turn to {\it null} circular geodesics in a Schwarzschild spacetime. For these orbits, unlike those considered in Fig.~\ref{fig:nearLR}, the sum over all multipole numbers $\ell$ does not appear to converge, and therefore we cannot compute the total  gravitational energy flux (either  at infinity or at the horizon). The contribution $\dot{E}_\ell$ of each multipole number $\ell$ seems to scale as $\dot{E}_\ell\propto 1/\ell$ for $\ell\gg 1$. This yields a logarithmically divergent $\dot{E}_{\rm tot}$, as suggested by the analytic results of Refs.~\cite{Misner:1972jf,PhysRevD.7.1002} (see also \cite{Detweiler:1978ge}). 

\begin{figure*}[t]
\begin{tabular}{cc}
\includegraphics[scale=0.35]{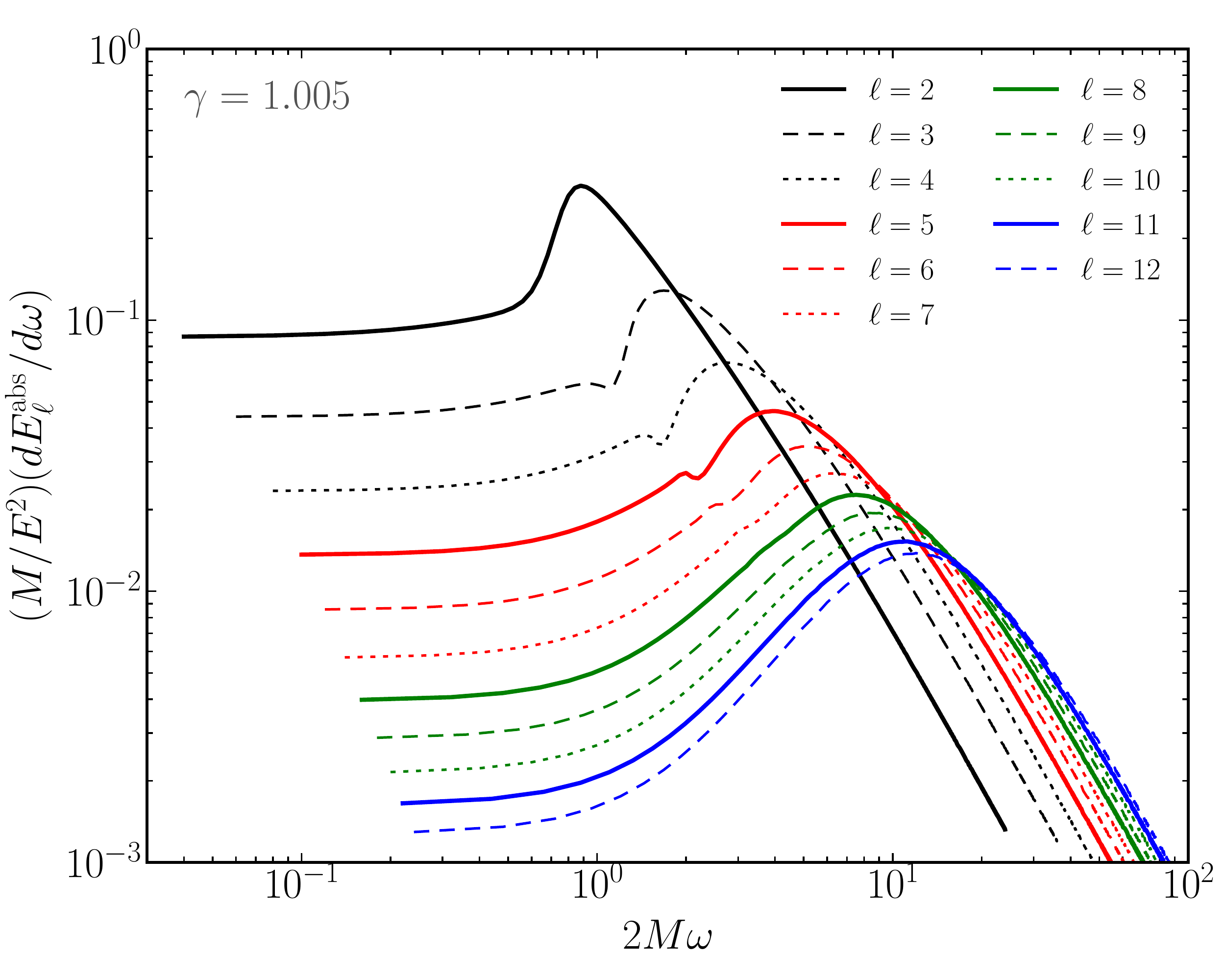}
\includegraphics[scale=0.35]{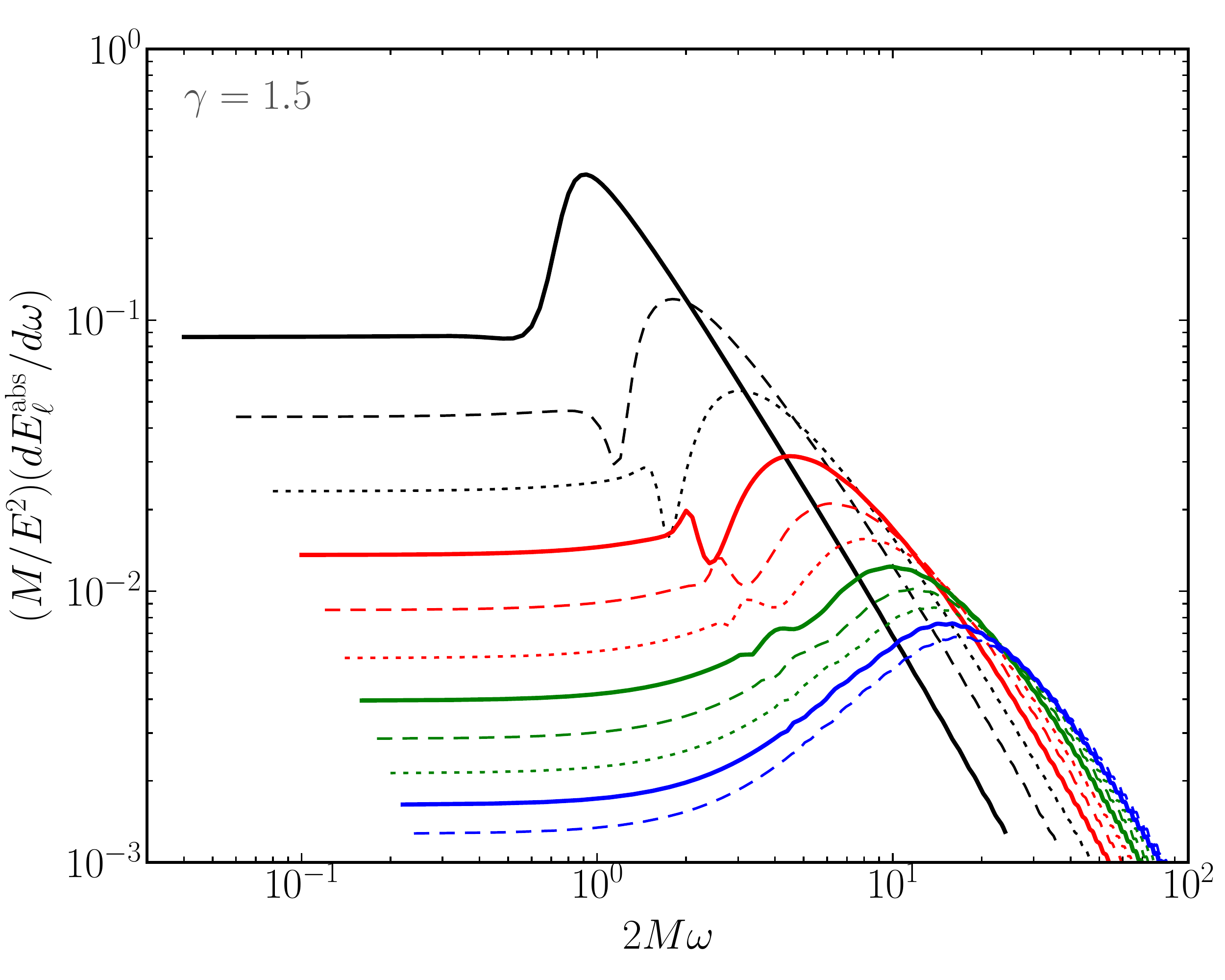}\\
\includegraphics[scale=0.35]{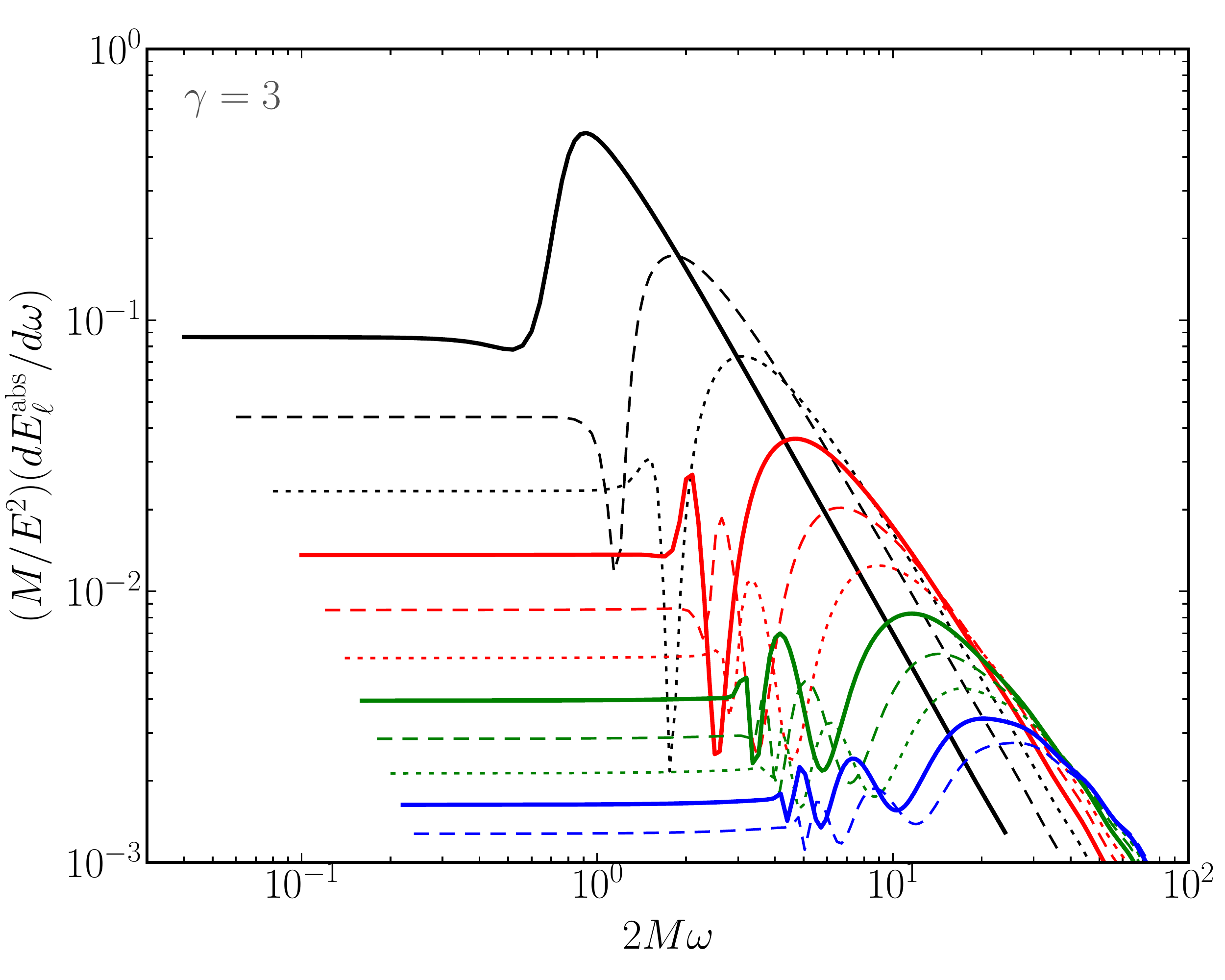}
\includegraphics[scale=0.35]{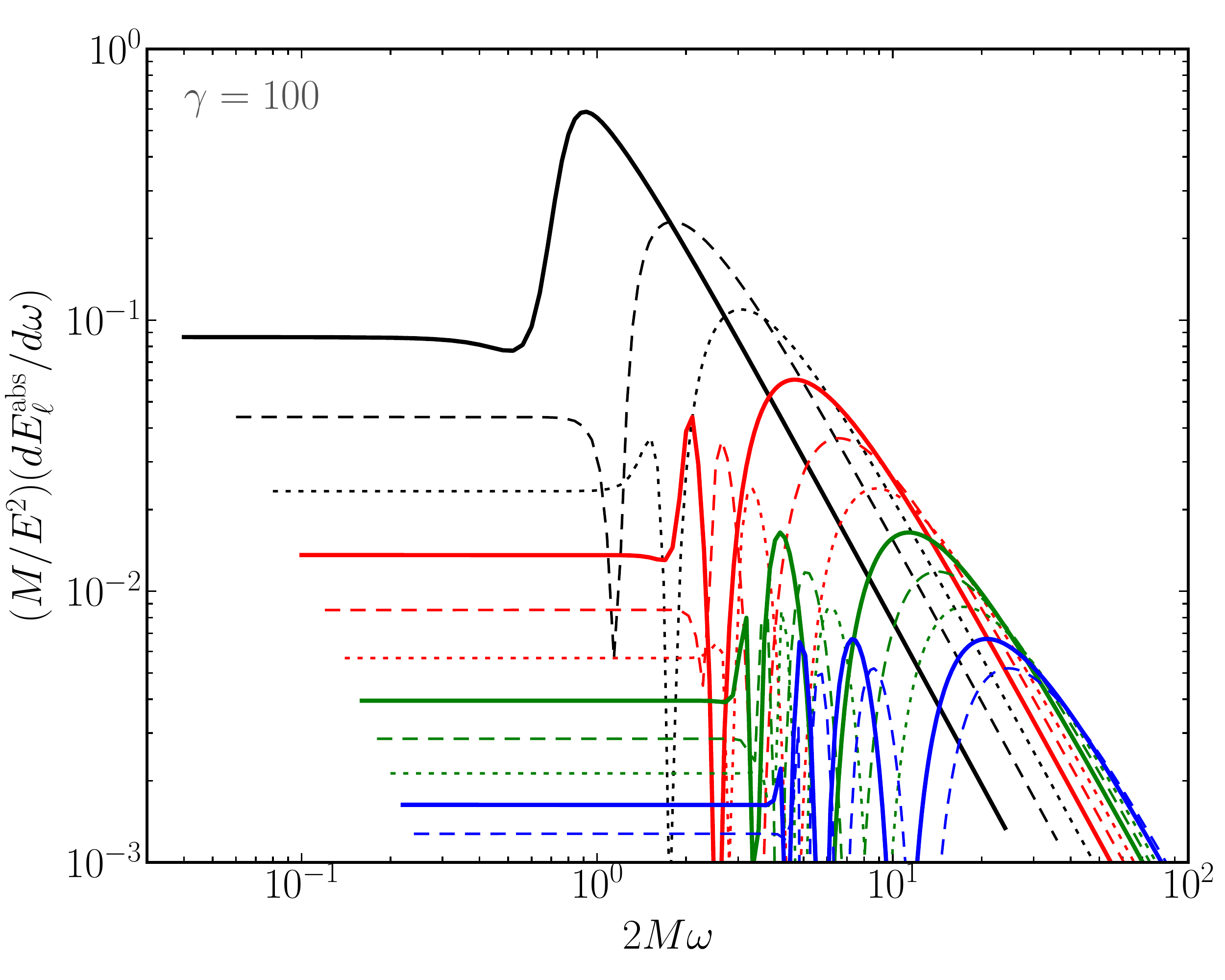}
\end{tabular}
\caption{Energy spectrum of gravitational radiation absorbed by a BH, in a given multipole $\ell$, as a function of $\ell$ when a particle plunges with Lorentz factor $\gamma=1.005, 1.5, 3, 100$.
}
\label{fig:plungedEdw}
\end{figure*}

\begin{figure}[t]
\includegraphics[scale=0.33,clip=true]{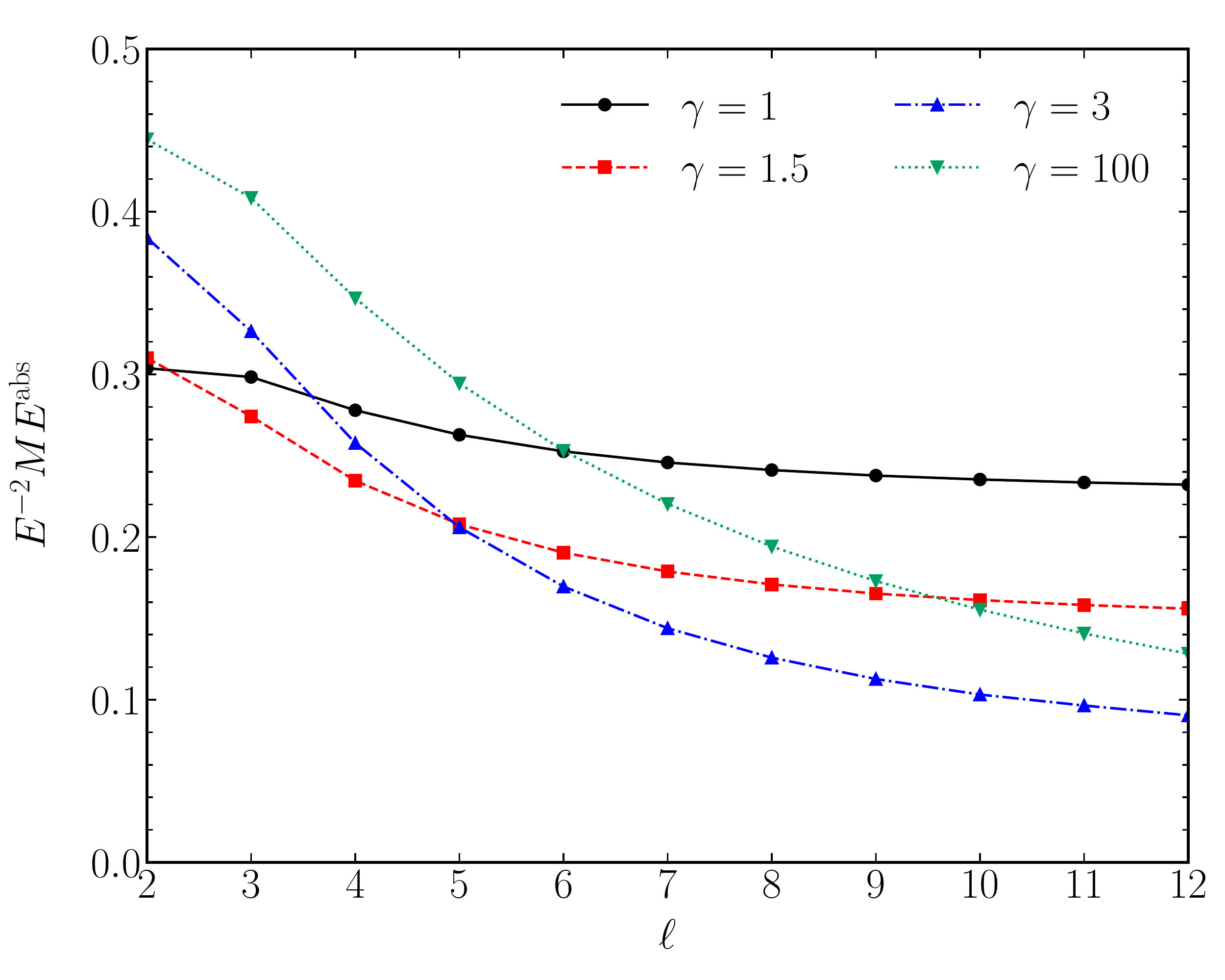}
\caption{Energy $E_\ell$ absorbed by the BH in a given multipole $\ell$, as a function of $\ell$ for particle energies $E=1.005$ (red), $E=1.5$ (green), $E=3$ (blue), $E=100$ (magenta).}
\label{fig:El}
\end{figure}

To verify the existence of such a logarithmic divergence, in the left panel of Fig.~\ref{fig:ratio} we show the difference between the ratio $\dot{E}_\ell/\dot{E}_{\ell-1}$ and the value [$(\ell-1)/\ell$] that would be expected from a $1/\ell$ scaling.  In the right panel of Fig.~\ref{fig:ratio} we show the difference between the ratio $\Delta \ln\dot{E}_\ell/\Delta \ln\ell$, with $\Delta \ln\dot{E}_\ell\equiv \ln\dot{E}_{\ell+1}-\ln\dot{E}_\ell$ and $\Delta \ln\ell\equiv \ln(\ell+1)-\ln\ell$, and the value $\Delta \ln\dot{E}/\Delta \ln\ell=-1$ that would be expected from the same $1/\ell$ scaling. These plots indicate that at large $\ell$ the fluxes do indeed approach the scaling $\dot{E}_\ell\propto 1/\ell$ predicted by WKB calculations~\cite{Misner:1972jf,PhysRevD.7.1002,ruffini}, but this behavior is not yet apparent at the multipole numbers $\ell\sim 50$ considered in Fig.~2 of Ref.~\cite{Davis:1972dm}, which are therefore not in disagreement with our results and those in Refs.~\cite{Misner:1972jf,PhysRevD.7.1002}. 
Our numerical results  at large $\ell$ are well fitted by 
\be
\dot{E}_\ell\approx \kappa \frac{E^2}{M^2\ell}\,,
\ee
with $\kappa\approx 0.064\pm0.001$. The scaling and proportionality constant $\kappa$ are the same for fluxes at infinity and at the horizon. This aspect is consistent with  the properties of synchrotron radiation from relativistic particles. Radiation is strongly forward-beamed at large Lorentz factors.
At the light ring the absorption angle is precisely 50\%~\cite{Shapiro:1983du,Cardoso:2019rvt}, thus giving rise to equal outgoing and ingoing fluxes.

Note also that the null geodesic result is consistent with its timelike counterpart. In the large-$\gamma$ limit, Eq.~\eqref{rel_particles} implies that the flux from relativistic timelike particles scales as $1/\ell$. Reassuringly our numerical fits yield $c_0\sim \kappa$, as they should if there is a smooth limit between timelike and null geodesics. We have extended this study to Kerr black holes and we find the same qualitative features.

\subsection{Energy absorbed by the horizon in head-on encounters}
\label{sec:hor}

We have seen above that both outgoing and ingoing instantaneous fluxes at the light ring diverge for particles on circular orbits. Another interesting divergence reported in the literature concerns radially infalling particles.
Consider a particle of energy $E=\gamma m_p$ plunging into a BH, which for definiteness we take to be nonspinning (the overall qualitative conclusions are general). The energy spectra in various multipoles per unit frequency bin and the total integrated energy in each multipole $\ell$ are shown for different infalling energies $\gamma$ in Figs.~\ref{fig:plungedEdw} and \ref{fig:El}, respectively.
These spectra were computed with minor modifications of the codes described in Refs.~\cite{Berti:2003si,Berti:2010gx}, and they refer exclusively to the energy fluxes going {\it into} the horizon.

The radiation spectra peak at $M\omega \sim \ell$, and then they decay as a power law. The energy spectra at large frequencies are well fitted by
\be
\frac{dE^{\rm abs}_\ell}{d\omega} \sim a_1 \gamma^2m_p^2(M\omega)^{-a_2}\,,
\ee
with $a_1,\,a_2$ constants which are only weakly dependent on the boost $\gamma$. We used the last 40 data points with higher frequency to fit the data.
For $\ell=2$, we find $(a_1,a_2)=(0.166, 1.944)$ for an energy $E=1.005$ and
$(a_1,a_2)=(0.197, 2.008)$ for an energy $E=100$. In the Appendix we develop a simple toy model which predicts $a_2=2$, in very good agreement with our data.

We get the total integrated energy in each multipole using the numerically computed values, complemented by this extrapolation when necessary. The multipolar contributions $E^{\rm abs}_\ell$ to the energy absorbed at the horizon obtained in this way are shown in Fig.~\ref{fig:El} (see~\cite{Berti:2003si,Berti:2010gx} for the spectra radiated at infinity, and Ref.~\cite{Berti:2010ce} for a simple analytical model and a comparison with ultrarelativistic simulations of head-on collisions in the comparable-mass limit~\cite{Sperhake:2008ga,Sperhake:2012me,Sperhake:2015siy}, where accretion also plays an important role~\cite{Sperhake:2009jz}). The total integrated energy going into the horizon is well described by expressions of the form
\be
\frac{M E^{\rm abs}_\ell}{E^2}=\kappa_0+\kappa_1 \ell^{-\kappa_2}\,.\label{behavior_multipolar_energy}
\ee
At low energies, $\kappa_2\sim 2$. For example, for $\gamma\sim 1$ we use the last 5 points of our data to find $\kappa_0=0.23, \kappa_1=0.99, \kappa_2=1.97$. These results are consistent with Ref.~\cite{Davis:1972ud}, where the total energy per multipole for infalls from rest is found to be roughly constant and well approximated by $\sim 0.25 m_p^2/M$.
For relativistic collisions, we find that $\kappa_0\sim0, \kappa_1 \sim 1.7,\, \kappa_2\sim 1$ at large $\ell$.

These results are very clear: $\kappa_0\neq 0$ and the total energy (summed over all multipoles with different values of $\ell$) diverges. We have studied also infalls in higher dimensional spacetimes~\cite{Cardoso:2002pa,Berti:2003si,Cook:2017fec}
(in particular spacetime dimensions $D=4,...,11$), in spacetimes with different asymptotics (in particular three-dimensional asymptotically anti-de Sitter BH spacetimes), and emission in different channels (most notably scalar radiation). While the qualitative behavior of the frequency spectrum and the multipolar content of the radiation vary, we consistently find that the total energy going into the BH horizon diverges in all of these cases. For the particular scenario of head-on infalls of point particles with higher dimensional non-spinning BHs~\cite{Cardoso:2002pa,Berti:2003si,Cook:2017fec}, for example, the divergence is even more pronounced for larger dimensionality.  We find a behavior similar to Eq.~\eqref{behavior_multipolar_energy}, but with $\kappa_2<0$. For spacetime dimension $D=6,7$, for example, our results indicate $\kappa_2\sim -1.4,-2.2$, thus decreasing very fast with $D$. 

\section{The role of regularization \label{sec:regularization}}

The divergence of the total radiation going into the horizon was already observed in Ref.~\cite{Davis:1972ud}. The authors intepreted the divergence as a consequence of the pointlike nature of the infalling object, which introduces high frequencies in the problem, and proposed to cure it by introducing a cutoff in the angular momentum expansion $\ell_{\rm crit} \sim \pi M/L$, where $L$ is the size of the particle: for example, $L\sim 2 m_p$ if the infalling particle is a BH.

The above procedure is somewhat ad hoc, and the cutoff needs to be properly justified. Below, we regularize the total fluxes by assuming that the orbiting object is a collection of ``dustlike'' particles, each of them pointlike. Such a procedure introduces explicitly a finite size for the orbiting objects, while keeping the calculation free of assumptions and mathematically correct at all stages. This regularization procedure leads to finite fluxes, providing further support to the argument that the pointlike nature of the objects is indeed the cause of the divergence.

\subsection{Circular motion}
\begin{figure}[t]
\includegraphics[width=0.5\textwidth]{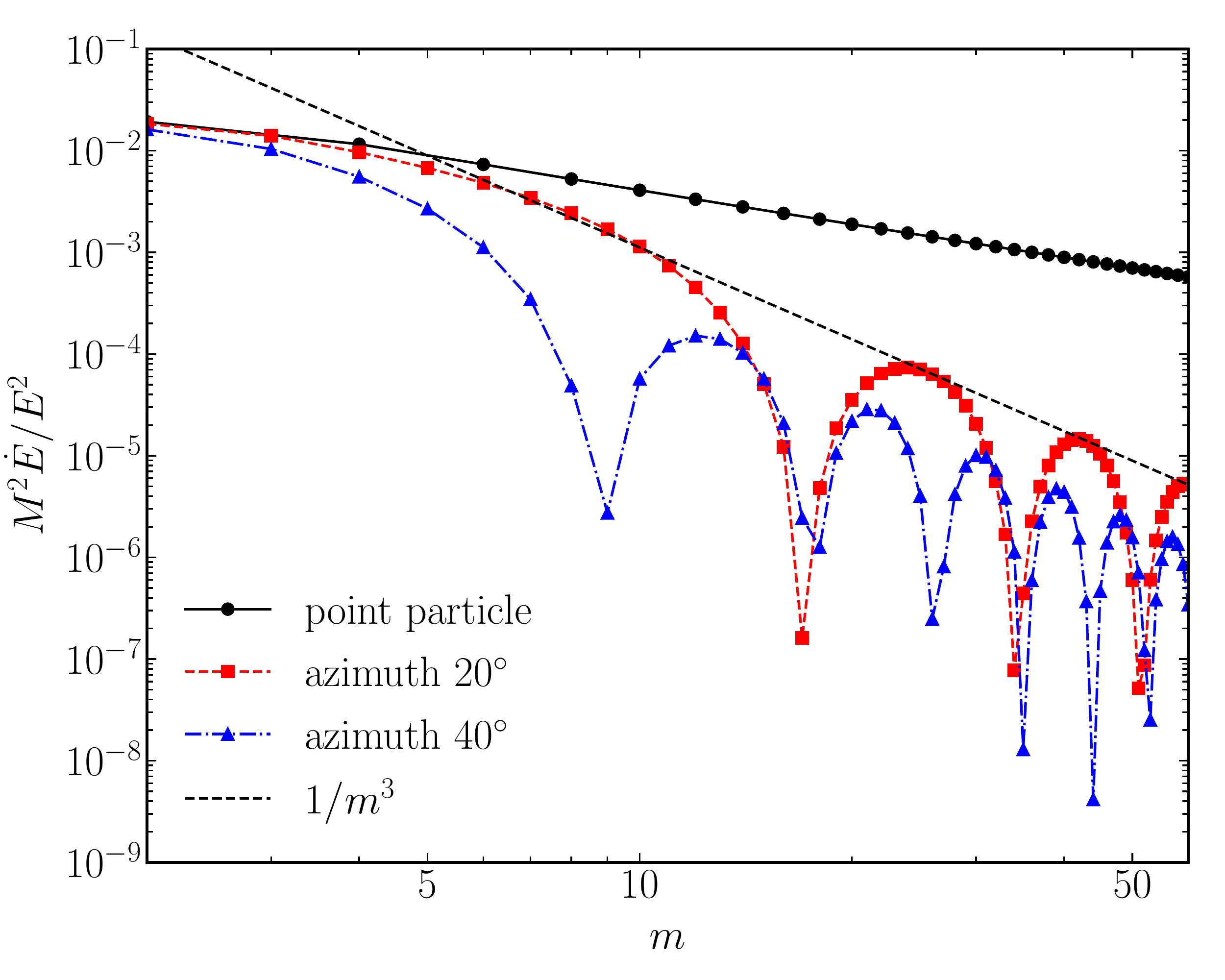} 
\caption{Dependence of $\ell=m$ fluxes at null infinity as a function of the azimuthal mode number $m$ from photons of varying sizes at the light ring of a Schwarzschild BH.
  For reference, we also overplot a straight line $\propto 1/m^3$. The oscillatory lines shows the fluxes in the dominant $\ell=m$ mode, when the particle is given a finite size in the azimuthal direction $\phi$. As expected from the superposition analysis leading to Eq.~\eqref{super_circular}, the flux has an oscillatory pattern of period inversely proportional to the size of the particle, and it decreases in amplitude as $\sim 1/m^3$.
}
\label{fig:eflux_width}
\end{figure}
For a single point particle in circular motion, the gauge-invariant metric fluctuations depend on the point particle mass as follows (see e.g.~\cite{Detweiler:1978ge}):
\be
h_++ih_\times=m_pe^{-i\omega (t-r_*)+im\phi}S_{\ell m}(\theta)Z_{\ell m}(\omega)\,.
\ee

Consider now $N$ pointlike particles, each of energy $E/N$, distributed in the azimuthal direction over an arc of length $L$ at radius $r_0$. In other words, the particles are located at $r = r_0$, $\phi = \phi_j$, where
\beq
\phi_j&=&\Omega t+\delta_j\,,\\
\delta_j&=&\frac{L}{r_0}\frac{j-1}{N}\,.
\eeq
We assume in addition that these are noninteracting ``dust'' particles, and since they are all located along the equator, they follow similar geodesics.
In our perturbative framework the equations are linear, so the single-particle wavefunctions add linearly. Given that the response due to each particle is dephased in the angle $\phi$ by $\delta_j$, we find that the total flux
\be
\dot{E}_{\rm tot}=\sum_{\ell m}\dot{E}_{1}|f(m,N)|^2\,,\label{super_circular}
\ee
where $\dot{E}_1$ is the $(\ell,\, m)$ component of the energy flux from a single pointlike particle of the {\it same total mass-energy $E$}, and
\be
f(m,N)=\sum_{j=1}^N\frac{e^{im\delta_j}}{N}\,.
\ee
In the large-$N$ limit we have
\be
|f(m,\infty)|^2=2(1-\cos{\eta})/\eta^2\,,\quad \eta\equiv 2\pi m L/r_0\,.
\ee
These results are consistent with previous findings~\cite{Nakamura:1981cd,Nakamura:1987zz,Haugan:1982fb}. At small $\eta$, $|f|^2=1$, thus the energy in the lowest multipoles is the same as that for a point particle. However,
at large values of $m$ we have $f \sim 1/m^2$. This is enough to cure all divergences. The transition occurs at $\eta \sim 1$, in rough agreement with our initial remarks.

The flux for a single point particle at the light ring behaves as $dE_{\ell m}/dt\sim 1/m$ at large $\ell=m$. This analysis (which is exact, within a frequency domain approach) also predicts that, for an extended particle along the azimuth, the flux scales like $\sim 1/m^3$, yielding a finite total energy flux.
These predictions are in very good agreement with numerical integrations of the Teukolsky equation with an extended source term in the time domain. In Fig.~\ref{fig:eflux_width} we show the total flux for each $\ell=m$ mode for a particle of finite size along the azimuthal direction. The particle was made finite by allowing for the source term to be nonzero over a range of $\phi$ values. We show results for extended sources of two different angular sizes, of $20^\circ$ and $40^\circ$, respectively. The exact analysis based on a collection of pointlike dust particles predicts an oscillatory flux with period $\propto 1/L$, and an envelope decaying as $1/m^3$. Both features are clear in our numerical results.
The divergence is indeed caused by high-frequency contributions which cancel out when superposing the signal from several point-particle sources with $\omega \gtrsim 1/L$ through interference effects, leading to a regularized total flux.

This frequency cutoff can be converted into a cutoff for the multipole number $\ell$: $\ell_{\rm max}\approx \pi/\Delta \theta$, where $\Delta\theta\sim L/(2 \pi r)$ and $r$ is the orbital radius. To understand the orders of magnitude involved, let us consider the radiation from M87$^{*}$ observed by the Event Horizon Telescope~\cite{Akiyama:2019cqa}. These waves have wavelength $L\sim $~mm, they are strongly lensed by the BH geometry, and at least some of them have orbited the light ring multiple times. Since the light ring radius $r=3 M$ (we assume a nonspinning BH for simplicity), this yields an enormous cutoff multipole number for M87$^{*}$, $\ell_{\rm max}\sim 10^{18}$. Nevertheless, summing up the GW fluxes plotted in Fig.~\ref{fig:nearLR} up to $\ell_{\rm max}$, we still find that the \textit{fractional} change of energy of a microwave photon at the light ring only amounts to $\sim 10^{-78}$ per cycle.

\subsection{Head-on encounters}
We can follow a similar strategy to regularize the total radiated energy going down the horizon for radial infalls. Analytical tools developed to understand radiation suppression at infinity for extended bodies~\cite{Nakamura:1981cd,Nakamura:1987zz,Haugan:1982fb,Berti:2006hb} can be used here as well. Consider first, for simplicity, a body extended along the radial direction~\cite{Berti:2006hb}. The normalized (Zerilli or Teukolsky) wavefunction
of a pointlike particle falling radially into a BH is $\Psi_1(t,r)$, or $\tilde{\Psi}_1(\omega, r)$ in the frequency domain, where by ``normalized'' we mean that the total field $\Psi(t,r)=m_p\Psi_1(t,r)$. Let us suppose that we now drop two such particles, each of mass $m_p/2$, which are separated by $L$ at large distances. Then we have
\be
\Psi(t,r)=\frac{m_p}{2}\left[\Psi_1(t,r)+\Psi_1(t-L,r)\right]\,,
\ee
with Fourier transform
\be
\tilde{\Psi}(\omega,r)=\frac{m_p}{2}\left[\tilde{\Psi}_1(\omega,r)+e^{i\omega L}\tilde{\Psi}_1(w,r)\right]\,.
\ee
The generalization to $N$ such bodies, each separated by $L/N$ and with a total spatial extent of $L$, is trivial:
\beq
\Psi(t,r)&=&\sum_{j=0}^{N-1}\frac{m_p}{N} \Psi_1\left(t-\frac{j L}{N-1},\,r\right)\,,\\
\tilde{\Psi}(\omega,r)&=&\sum_{j=0}^{N-1}\frac{m_p}{N}e^{i\frac{\omega L}{N-1}j}\tilde{\Psi}_1(\omega,\,r)\,.
\eeq
When $N\to \infty$ the summation can be done analytically with the result
\be
\tilde{\Psi}(\omega,r)=\frac{-1+e^{i\omega L}}{i\omega L}\tilde{\Psi}_1(\omega,r)\,.
\ee
and the energy spectrum reads
\beq
\frac{dE}{d\omega}&=&
g(\omega)\frac{dE_1}{d\omega}\,,\\
g(\omega)&=&\frac{2-2\cos\omega L}{\omega^2L^2}\,.
\eeq
As before, at small frequencies ($\omega\sim 0$) we have $g(\omega)\sim 1$. However, at large frequencies (which is also equivalent to large $\ell$, see the discussion around Fig.~\ref{fig:plungedEdw}) one gets $g \sim 1/\omega^2$. Thus the divergence is regularized, and the procedure is equivalent to introducing a cutoff frequency $\omega \sim 1/L$, which is equivalent to a cutoff multipolar index $\ell_{\rm crit}\sim M/L$, as argued on heuristic grounds in Ref.~\cite{Davis:1972ud}.

Using this regularization scheme for small particles, with $\ell_{\rm crit}\sim M/(2m_p)\gg 1$, we find that an infall from rest absorbs
\be
\frac{E^{\rm abs}}{m_p}\sim \frac{m_p}{M}(2.8+0.23\ell_{\rm cut})\sim 0.1 m_p\,.
\ee
Thus the absorbed energy is of the order of $10\%$ of the rest mass of the infalling particle. 

For relativistic collisions of point particles we find
\be
E^{\rm abs}\sim \frac{E^2}{M}(2.7+\log\ell_{\rm cut})\,,
\ee
hence the absorbed energy is now a small fraction of the energy of the incident point particle.

For nearly equal-mass collisions -- which require extrapolation of our results -- we can use only the $\ell=2$ mode. This cutoff is consistent with both the above suggestion and with numerical simulations for the total radiated energy at infinity. For low-energy infalls, the energy of the small particle, $E$, can be promoted to the reduced mass of the system, while the BH mass is promoted to the total mass~\cite{Anninos:1995vf,Cardoso:2002ay,Witek:2010az}. 
We then find a total absorbed energy of the order of $2\%$ the total mass of the system.

\section{Conclusions}
For most applications in GW physics, point particle approximations have served us well. The results are convergent, and radiation at infinity is dominated by wavelengths comparable to the size of the system. When extrapolated to equal-mass processes the results are, in addition, surprisingly accurate~\cite{Berti:2007fi,Berti:2010ce,Tiec:2014lba,Rifat:2020}.

However, this approach fails in two circumstances at least. One is in the calculation of radiation from a null particle on the (unstable) null geodesic.
The radiation has contribution from arbitrarily large frequencies and it formally diverges. The divergence is ``mild'' when expressed in multipolar components. 
We have shown that this divergence can be cured by assigning a finite size to the null particle, leading to finite radiation fluxes, and to equal fluxes at the horizon and infinity. As a side result, we have also shown that the backreaction on null geodesics is never significant enough to affect any of the physics related to the light ring (for example in the context of the Event Horizon Telescope).

The second circumstance where the point particle approach fails is in the calculation of radiation going down the hole, say during the collision of two BHs. A multipolar decomposition of the radiation implies that the gravitational waves going down the BH carry an important (in fact divergent) component of the energy in high-frequency waves. This is surprising in light of the fact that the burst at infinity is dominated by waves of frequency $\omega \sim M$ and by the lowest multipoles. We do not have an elegant interpretation for such a divergence/convergence duality, but we note that the divergent energy at the horizon is naturally cured by introducing finite-size effects for the infalling particle. In other words, our calculations show that the point particle divergences are always ``soft'' enough that they are easily cured with finite-size effects.

The problem of motion in general relativity seems to satisfy an ``effacement principle'' (see e.g.~\cite{Damour:1986ny}), in the sense that the physics is never too sensitive to small scales. It seems after all that this property fails to apply in the two examples we discussed: small scale structure and the composition of the two bodies is important to determine the leading-order radiation effects. A broader understanding of this aspect would certainly merit consideration.
These examples all involve particles traveling at or approaching the speed of light. Relativistic beaming thus seems responsible for the ensuing divergences, although we do not have a formal generic proof.

\noindent{\bf{\em Acknowledgments.}}
E.~Barausse acknowledges financial support provided under the European Union's H2020 ERC Consolidator Grant ``GRavity from Astrophysical to Microscopic Scales'' grant agreement no. GRAMS-815673.
E.~Berti is supported by NSF Grants No. PHY-1912550 and AST-2006538, NASA ATP Grants No. 17-ATP17-0225 and 19-ATP19-0051, NSF-XSEDE Grant No. PHY-090003, and NSF Grant PHY-20043. This research project was conducted using computational resources at the Maryland Advanced Research Computing Center (MARCC).
V.~Cardoso acknowledges financial support provided under the European Union's H2020 ERC 
Consolidator Grant ``Matter and strong-field gravity: New frontiers in Einstein's 
theory'' grant agreement no. MaGRaTh--646597.
S.~A.~Hughes is supported by NSF Grant No.~PHY-1707549 and NASA ATP Grant 80NSSC18K1091.
G.~Khanna acknowledges research support from NSF Grants PHY-2106755 \& DMS-1912716 and ONR/DURIP Grant No. N00014181255.
This project has received funding from the European Union's Horizon 2020 research and innovation programme under the Marie Sklodowska-Curie grant agreement No 101007855.
We thank FCT for financial support through Project~No.~UIDB/00099/2020.
We acknowledge financial support provided by FCT/Portugal through grants PTDC/MAT-APL/30043/2017 and PTDC/FIS-AST/7002/2020.
The authors would like to acknowledge networking support by the GWverse COST Action 
CA16104, ``Black holes, gravitational waves and fundamental physics.''
This work makes use of the Black Hole Perturbation Toolkit.

\appendix
\section{A model problem: scalar emission from a plunging particle}

We are interested in understanding the features of gravitational radiation going {\it into}
the BH horizon as a point particle plunges in. Let us consider for simplicity scalar particles.
At first order in perturbation theory, the field equations for the scalar field reduce to 
\be
\left[\square-\mu_s^2\right]\varphi=\alpha {\cal T}\label{EQBD2b}\,,
\ee
where $\mu_s$ is the scalar field mass, which we will take to be zero, and $\alpha$ the (scalar) charge of the infalling 
particle. Since we are describing a point particle, we focus on source terms of the form
\be
{\cal T}=\int \frac{d\bar\tau}{\sqrt{-\bar g^{(0)}}}\,m_p\delta^{(4)}\left(x-X(\bar\tau)\right)\,,
\ee
corresponding to the trace of the stress-energy tensor of a point particle with mass $m_p$, where $\bar g^{(0)}$ is the determinant of the background metric. In scalar-tensor theories, for example, $\alpha=\sqrt{{8 \pi}/(2+\omega_{\text{BD}})} \left(s-1/2\right)$, where $\omega_{\rm BD}$ is the Brans-Dicke (BD) parameter and $s$ is an object-dependent ``sensitivity'' factor.%

Because of the coupling to matter, the object emits both gravitational and scalar radiation. Here we focus on scalar wave emission.
We will also focus on Schwarzschild backgrounds, although our results are easily generalized. Consider a particle falling along the axis of a Schwarzschild BH, and decompose the scalar field as
\be
\varphi(t,r,\theta)=\sum_{\ell}\int d\omega e^{-i\omega t}\frac{X_{\ell}(\omega,r)}{r\sqrt{2\pi}} Y_{\ell 0}(\theta)\,,
\ee
where $Y_{\ell 0}$ are scalar spherical harmonics.
We get the inhomogeneous equation for the scalar field
\be
\left[\frac{d^2}{dr_*^2}+\omega^2-V\right]X_{\ell \omega}(r)=f\,T_{\ell \omega}\,,\label{nonhom0}
\ee
where $dr/dr_*=f\equiv 1-2M/r$,
\beq
T_{\ell \omega}&=&-\frac{\alpha m_p}{\sqrt{2\pi}\,r}Y^*_{00}(0)\,e^{i\omega T(r)}\left(dr/d\tau\right) ^{-1}\label{tlmw}\,,\\
V&=&f\left[\frac{\ell(\ell+1)}{r^2}+\frac{2M}{r^3}\right]\,.
\eeq

For a particle falling straight in, geodesic motion requires that
\be
\frac{dT}{dr}=-\frac{E}{f\sqrt{E^2-f}}\,,\qquad \frac{dT}{d\tau}=\frac{E}{f}\,,
\ee
where $E$ is a conserved energy parameter.
Let us consider two independent solutions $X_{\ell \omega}^{r_+}$ and $X_{\ell \omega}^{\infty}$ to the homogeneous equation satisfying the following boundary conditions:
\be
X_{\ell \omega}^{\infty,r_+}\sim e^{\pm i \omega r_*} \quad{\rm as} \quad r_*\to \pm\infty\,,
\ee
and let $W$ be their Wronskian.
The spectrum of scalar energy at the horizon and at infinity reads
\beq
\frac{dE^{s}_{r_+,\infty}}{d\omega}&=&\omega^2|Z^{r_+,\infty}|^2\,,\label{scalar_fluxes}\\
Z_{\ell \omega}^{r_+,\infty}&\equiv&\int_{r_+}^{\infty}\frac{T_{\ell \omega}\, X_{\ell \omega}^{r_+,\infty}}{W}\,.
\eeq

To summarize, the energy spectrum is determined in two steps: 
one first finds a solution of the homogeneous ODE and then integrates it over the source term.
In the high-frequency regime, the first step becomes trivial. The ODE is just a harmonic oscillator
and $X_{\ell \omega}^{\infty,r_+}=e^{\pm i \omega r_*}$. The Wronskian is then $W=2i\omega$.
Thus, the only thing now left to do is to perform the integral
\be
\int_{r_+}^{\infty}\frac{\left(dr/d\tau\right)^{-1}}{r}\frac{e^{i\omega \left(T(r)\pm r_*\right)}}{2i\omega}dr\,.
\ee
The integral, and therefore the spectrum, is highly dependent on
the asymptotic properties of the phase of the exponential: $T(r)-r_*$ goes to a constant at the horizon,
whereas $T(r)+r_*$ diverges there. As such one expects that the integral averages out more rapidly at higher frequencies
when computing fluxes at infinity. In other words, one expects that the high frequency fall-off of fluxes at infinity
is faster than for fluxes at the horizon.

To see this, let's approximate the integrand.
The coordinate $r_*=r+2M\log(r-2M)$, and we approximate
$T(r)\simeq r -2M\log(r-2M)$, while keeping $dr/d\tau={\rm const}$. Notice that asymptotically $T\to \infty$ at both the horizon and infinity, but this is not a problem, as one can always redefine a new time coordinate and split the integration interval into two.
In any case one gets exact expressions:
\begin{widetext}
\beq
&\int_{r_+}^{\infty}\frac{e^{i\omega \left(T(r)- r_*\right)}}{r}dr=\Gamma(0,-4iM\omega)\sim(-4iM\omega)^{-1}e^{4iM\omega}\,,\nonumber\\
&\int_{r_+}^{\infty}\frac{e^{i\omega \left(T(r)+ r_*\right)}}{r}dr=-i(2M)^{-4 I M \omega} \pi {\rm cosech}(4\pi M \omega)\sim e^{-4\pi M \omega}\,.\nonumber
\eeq
\end{widetext}
This simple exercise predicts that the spectrum of energy going into the hole decays as $1/\omega^2$, a result which can also be obtained rigorously with an integration by parts~\cite{Huybrechs}. As we mentioned in the main text, our numerical results are in very good agreement with this expected behavior.

This analysis also predicts that the flux at infinity scales like $e^{-8\pi M\omega}$,
a prediction which has been obtained by Zerilli with a steepest-descent analysis~\cite{Zerilli:1971wd}. If we fit our data for the spectrum at infinity to an exponential decay $\sim e^{-a_3\omega}$, we find for $\ell=2$ 
that $a_3=(26.51,\,28.42)$ for $E=1.005,\,100$ respectively, still in good agreement with the $8\pi$ prediction. Our results for higher-dimensional spacetimes can be interpreted with the same toy model with $f=1-r_H^{D-3}/r^{D-3}$, where $r_H$ is the horizon radius in Schwarzschild-like coordinates and $D$ the spacetime dimension. The absorbed radiation spectrum still scales like $\omega^{-2}$ at large frequencies in both this toy model and in our numerical results.

\bibliographystyle{apsrev4-1}
\bibliography{References}

\end{document}